\pdfoutput=1
\documentclass[11pt,letterpaper]{article}

\usepackage[utf8]{inputenc}
\usepackage[T1]{fontenc}
\usepackage{newtxtext,newtxmath}
\usepackage[letterpaper,margin=1in,footskip=0.45in]{geometry}
\usepackage{microtype}
\usepackage{setspace}
\usepackage{graphicx}
\usepackage{booktabs,array}
\usepackage[font=small,labelfont=bf,labelsep=period,justification=justified,singlelinecheck=false]{caption}
\usepackage{enumitem}
\usepackage{needspace}
\usepackage[authoryear,round,sort]{natbib}
\usepackage{authblk}
\usepackage{titlesec}
\usepackage{fancyhdr}
\usepackage{xcolor}
\usepackage{tikz}
\usetikzlibrary{arrows.meta,positioning,calc}
\definecolor{navy}{RGB}{20,45,100}
\definecolor{notegray}{RGB}{110,110,110}
\usepackage[colorlinks=true,linkcolor=navy,citecolor=navy,urlcolor=navy,hyperfootnotes=false,
            pdftitle={AI-Mediated Employee Voice and Performance: Disentangling Informational and Affective Pathways},
            pdfauthor={Thilo Tamme, Andrej Levin},
            pdfkeywords={employee voice, artificial intelligence, performance, well-being, kaizen, professional services, field experiment}]{hyperref}

\titleformat{\section}{\large\bfseries}{\thesection}{0.8em}{}
\titleformat{\subsection}{\normalsize\bfseries}{\thesubsection}{0.8em}{}
\titleformat{\subsubsection}[runin]{\normalsize\bfseries\itshape}{}{0pt}{}[.]
\titlespacing*{\section}{0pt}{2.2ex plus 1ex minus .2ex}{1.2ex plus .2ex}
\titlespacing*{\subsection}{0pt}{1.8ex plus .8ex minus .2ex}{0.8ex plus .2ex}
\titlespacing*{\subsubsection}{0pt}{1.2ex plus .5ex}{0.6em}
\newcommand{\SMSnote}{This research proposal was accepted, following double-blind peer review, for presentation at the
  46th Annual Conference of the Strategic Management Society (SMS), Berlin, Germany, October 17--20, 2026.}
\newcommand{\SMSfooter}{Accepted after double-blind peer review for the 46th SMS Annual Conference, Berlin, October 17--20, 2026}

\makeatletter
\newcommand{\acceptancenote}[1]{%
  \begingroup
  \renewcommand\thefootnote{}%
  \long\def\@makefntext##1{\noindent\footnotesize ##1}%
  \footnotetext{#1}%
  \endgroup}
\makeatother

\fancypagestyle{plain}{\fancyhf{}\fancyfoot[R]{\small\thepage}}

\newcommand{\hyp}[2]{\begin{quote}\textbf{#1}\ \textit{#2}\end{quote}}

\title{\vspace{-1.2em}{\normalsize\scshape Research Proposal}\\[0.9em]
  {\LARGE\bfseries AI-Mediated Employee Voice and Performance:}\\[0.3em]
  {\Large\bfseries Disentangling Informational and Affective Pathways}}

\author[1]{Thilo Tamme}
\author[1]{Andrej Levin}
\affil[1]{Technical University of Munich}
\date{February 2026}

\begin{document}

\maketitle
\thispagestyle{plain}
\acceptancenote{\SMSnote}

% ---------------------------------------------------------------------
\begin{abstract}
\noindent
Employee voice---the discretionary expression of ideas, concerns, and suggestions intended to improve organizational functioning---creates value inside firms, yet organizations increasingly replace or augment human-led listening with AI-enabled systems that collect, analyze, and route employee input at scale. Whether algorithmic listening replicates the value-creation mechanisms of human voice, and whether it alters the pathways through which voice translates into performance, remains an open question. This research proposal theorizes that traditional voice bundles two distinct functions, knowledge extraction and relational signaling, and that AI mediation may decouple them: strengthening the informational pathway (implemented improvement measures) while weakening the affective pathway (the experience of being heard). We propose a randomized field intervention at a global strategy consulting firm, covering approximately 2,000 client-facing consultants, that varies the form of listening across three conditions---no structured voice, human-led survey-based voice, and AI-enabled conversational voice---and links weekly team survey data to longitudinal performance ratings. A dual-mediation model is tested in which voice affects performance through (1) implemented process improvements and (2) affective well-being. The study extends voice theory to knowledge-intensive professional services, disentangles the micro-mechanisms of voice, and provides causal evidence on whether AI can serve as an effective listener.
\end{abstract}

\vspace{0.3em}

% ---------------------------------------------------------------------
\section{Introduction}

Employee voice---the discretionary expression of ideas, concerns, and suggestions intended to improve organizational functioning---has long been understood as a source of value creation inside firms \citep{farndale2011,morrison2011,nyfoudi2023,rees2013,wilkinson2020}. Recent evidence from a large-scale MIT study of 30,000 U.S.\ manufacturers suggests that worker voice can substantially increase firm productivity \citep{nelson2025}. Specifically, workplaces incorporating worker voice showed a 16\% rise in productivity surplus in manufacturing (Nelson, 2026).%
\ These findings reinforce a core strategic insight: organizations that effectively harness employee voice outperform those that do not.

Yet a fundamental question remains unresolved.

Voice research has historically assumed that listening is relational---embedded in human interaction, managerial responsiveness, and social exchange (Morrison, 2012).%
\ But organizations are increasingly replacing or augmenting human-led listening mechanisms with AI-enabled systems that collect, analyze, and route employee input at scale \citep{malik2023}. This raises a deeper strategic puzzle:

\begin{quote}
\itshape To what extent does AI-mediated listening replicate the value-creation mechanisms of human voice, and does AI mediation alter the pathways through which employee voice translates into performance?
\end{quote}

This question is not merely technological. It concerns the microfoundations of employee, team, and firm performance.

Prior research suggests that voice creates value through at least two mechanisms. First, voice enables process improvement by surfacing localized, match-specific knowledge that managers cannot directly observe, or improvement opportunities that they may fail to capture \citep{nelson2025}. When implemented, such employee input enhances efficiency and productivity, often referred to as kaizen measures in production contexts \citep{kasahara2026}. Second, voice produces affective benefits: the experience of ``being heard'' signals respect, psychological safety, and organizational support, thereby increasing employee commitment and effort \citep{li2017}. Both pathways have been theorized in general terms, yet they rest on different assumptions about how value is generated---one informational, one relational.

AI-enabled voice systems may affect these two mechanisms quite differently.

On the one hand, AI tools can aggregate dispersed knowledge more rapidly and consistently than human managers, potentially amplifying the informational benefits of voice, which we refer to as improvement measures (kaizen). On the other hand, AI mediation may weaken the relational signaling that underpins psychological safety and perceived support. Employees may interpret algorithmic listening as impersonal or even surveillance-oriented, thereby attenuating affective gains (Li et al., 2025).%
\ Thus, AI-mediated voice may strengthen process-based value creation while simultaneously weakening relational value creation---or vice versa. Whether AI substitutes for, complements, or fundamentally transforms traditional voice mechanisms remains theoretically and empirically unclear.

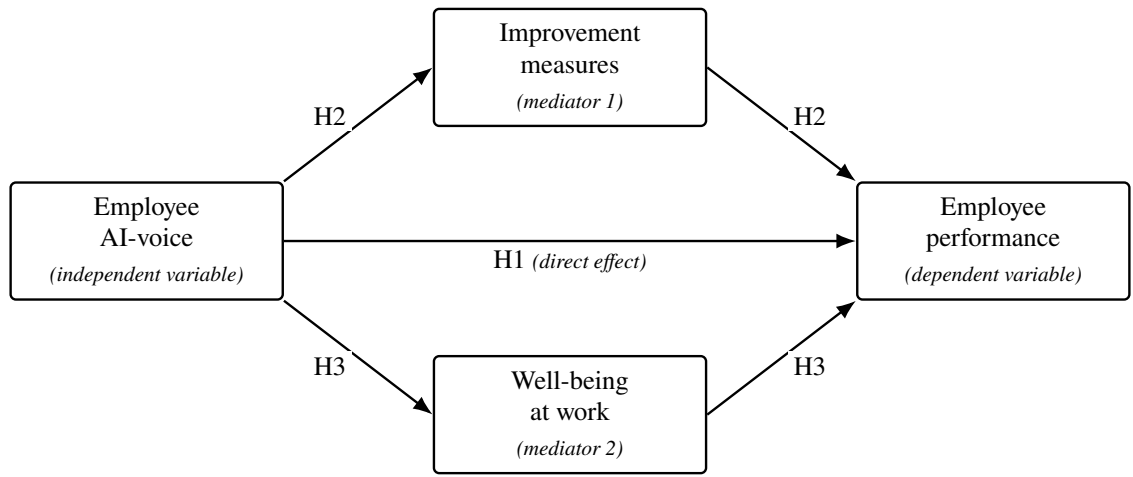
\begin{figure}[htbp]
\centering
\begin{tikzpicture}[
  box/.style={draw=black, line width=0.9pt, rounded corners=2pt, align=center,
              minimum width=3.6cm, minimum height=1.55cm, font=\small, fill=white},
  sub/.style={font=\scriptsize\itshape},
  arr/.style={-{Latex[length=2.6mm,width=2mm]}, line width=0.9pt},
  lab/.style={font=\small, fill=white, inner sep=2pt}
]
\node[box] (iv) at (0,0)      {Employee\\AI-voice\\[1pt]{\scriptsize\itshape (independent variable)}};
\node[box] (m1) at (5.6,2.3)  {Improvement\\measures\\[1pt]{\scriptsize\itshape (mediator 1)}};
\node[box] (m2) at (5.6,-2.3) {Well-being\\at work\\[1pt]{\scriptsize\itshape (mediator 2)}};
\node[box] (dv) at (11.2,0)   {Employee\\performance\\[1pt]{\scriptsize\itshape (dependent variable)}};
\draw[arr] (iv.north east) -- node[lab, pos=0.45, above left=-1pt] {H2} (m1.west);
\draw[arr] (m1.east) -- node[lab, pos=0.55, above right=-1pt] {H2} (dv.north west);
\draw[arr] (iv.east) -- node[lab, below=1pt] {H1 \scriptsize\itshape (direct effect)} (dv.west);
\draw[arr] (iv.south east) -- node[lab, pos=0.45, below left=-1pt] {H3} (m2.west);
\draw[arr] (m2.east) -- node[lab, pos=0.55, below right=-1pt] {H3} (dv.south west);
\end{tikzpicture}
\caption{Parallel mediation model: employee AI-voice and employee performance. H1 denotes the direct effect; H2 the indirect effect through implemented improvement measures (informational pathway); H3 the indirect effect through affective well-being (relational pathway).}
\label{fig:model}
\end{figure}

This study addresses this gap by examining employee voice within a global strategy consulting firm across approximately 2,000 client-facing consultants, leveraging a randomized intervention study that varies the form of listening: no structured voice; human-led, survey-based voice; and AI-enabled conversational voice. Using longitudinal performance and well-being data, we test a dual-mediation model in which voice affects employee performance through (1) implemented process improvements and (2) affective well-being (Figure~\ref{fig:model}).

By disentangling informational and relational pathways, and by comparing human versus AI-mediated listening, this study contributes to three conversations.

First, this study advances emerging research on AI in organizations by theorizing and empirically testing whether AI can function as an effective listener. While organizations are rapidly adopting AI-enabled voice platforms, we lack causal evidence on whether algorithmic listening can replicate or even outperform human-mediated listening mechanisms \citep{tambe2019}. We examine whether AI-mediated voice increases participation, improves the quality of improvement measures, and sustains (or undermines) well-being. By directly comparing AI-enabled conversational agents with human-led voice channels in a randomized field setting, the study moves beyond speculation and provides systematic evidence on whether AI constitutes a superior, complementary, or substitutive voice channel.

Second, our research disentangles the micro-mechanisms through which employee voice translates into performance advantages. Prior research implicitly assumes that informational and relational effects are aligned \citep{li2017,morrison2011,romney2024}. We theorize that they are structurally distinct and may diverge under algorithmic mediation. We develop and test a dual-mediation framework that distinguishes between (1) an informational pathway, in which voice generates implemented process improvements, and (2) an affective pathway, in which the experience of being heard enhances psychological well-being and, in turn, performance. By isolating and comparing these mechanisms, we clarify how voice creates value and assess whether AI-mediated listening differentially strengthens informational gains while potentially weakening relational signaling effects.

Third, this study extends voice theory to knowledge-intensive professional service firms, a context absent from existing empirical evidence. Most productivity-based evidence on employee voice originates from manufacturing and blue-collar settings \citep[e.g.,][]{nelson2025}, leaving open the question of generalizability to white-collar knowledge work. In professional services, expertise is tacit, performance is multidimensional, and contributions are difficult to observe or standardize. Moreover, measuring performance in such environments presents methodological challenges, as traditional output metrics often fail to capture nuanced value creation. By leveraging a clearly defined and longitudinal performance evaluation system within a global consulting firm, this study provides rare causal evidence on voice--performance dynamics in knowledge-intensive work and expands the boundary conditions of voice theory.

In an era where firms increasingly rely on algorithmic systems to mediate internal communication, understanding whether---and how---AI can serve as an effective listener represents a critical frontier for strategic management research.

% ---------------------------------------------------------------------
\section{Theoretical Background and Hypotheses}

Traditional employee voice mechanisms inherently bundle two distinct organizational functions. First, they serve as a knowledge extraction device: managers solicit and interpret frontline insights, surfacing dispersed, tacit, and match-specific knowledge that would otherwise remain localized. Second, they function as a relational signaling mechanism: the very act of being listened to by a human authority communicates respect, inclusion, and psychological safety.

In conventional settings, these two functions are intertwined because the same managerial interaction both extracts information and conveys relational value. AI-mediated listening, however, has the potential to decouple these functions.

Algorithmic systems may enhance the efficiency and scale of team insight aggregation---identifying patterns, prioritizing issues, and routing suggestions more consistently than human managers---while simultaneously weakening or altering the relational signal traditionally conveyed through interpersonal attention. As a result, AI may strengthen the informational pathway through which voice improves performance while attenuating, transforming, or even reversing the affective pathway rooted in human recognition. Such decoupling would reveal that employee voice is not a unitary construct but a bundled governance mechanism whose informational and relational components may respond differently to algorithmic mediation.

\subsection{Employee Voice and Performance: Direct Effects}

Employee voice encompasses both promotive voice (new ideas for improvement) and prohibitive voice (concerns about harmful practices) \citep{liang2012,romney2024}. Extant research demonstrates that voice behavior relates positively to organizational decision-making, employee engagement, trust in management, and organizational commitment \citep{farndale2011,rees2013}. Moreover, voice behavior has been linked to employees' creativity, innovative behavior, and other-rated in-role performance \citep{chen2020,ng2012}.

The value-creation logic underlying these relationships is straightforward: providing constructive ideas or raising concerns and grievances regarding employee-related or organizational problems may help employees, as well as organizations, weather crises (Wee \& Fehr, 2021; Wilkinson et al., 2020, as cited in Nyfoudi et al., 2023). As employees identify problems, organizations can prevent large-scale failures, avoid costly errors, and protect employee well-being \citep{edmondson2004,li2017}. Through voice, workers can ``effectively ask for shifts in the composition of their compensation and work'' and achieve ``win-win improvements in working conditions that maintain job quality while benefiting both workers and employers'' (Freeman \& Medoff, 1984, as cited in Nelson \& Wilmers, 2025, p.~4).

\hyp{H1.}{Employee AI-voice is positively associated with employee performance.}

\subsection{Process Mediation: The Improvement-Measure Effect}

The first mediating pathway linking voice to performance operates through process improvement measures. When employees share insights gleaned from daily work experience, organizations can identify and implement improvement measures that enhance operational efficiency and output quality. This mechanism reflects the traditional kaizen philosophy of continuous improvement through worker participation \citep{kasahara2026}.

Recent evidence supports this process-oriented pathway. Romney et al.\ \citeyearpar{romney2024} analyzed 555 employee voice episodes over four years and found that prohibitive voice---communication of problems impairing organizational effectiveness---was positively associated with response urgency, and urgent voice episodes were significantly more likely to be implemented (83.1\% vs.\ 71.1\% on average). Critically, ``the indirect effect of prohibitive voice was positively related to implementation through response urgency'' \citep[p.~18]{romney2024}, suggesting that voice triggers supervisory action that translates into tangible organizational improvements.

This implementation of worker suggestions creates match-specific surplus. As Nelson and Wilmers \citeyearpar{nelson2025} explain: ``A production worker's perceptive comments about process improvement, gleaned from years of experience at a given plant, reduce the cost of model changeovers. If that worker left the company, the employer would lose the match specific surplus gained from that worker's voice'' (p.~7). When these improvements are implemented, they increase employees' ability to perform their work more effectively, creating a positive feedback loop.

\hyp{H2.}{The relationship between employee AI-voice and employee performance is positively mediated by the number of improvement measures identified and implemented.}

\subsection{Affective Mediation: The Well-Being Effect}

The second mediating pathway operates through affective mechanisms, independent of whether suggestions are implemented. The experience of ``being heard''---of having one's input solicited, acknowledged, and valued---positively affects employee well-being and job satisfaction, which in turn enhances performance.

HR practices serve as ``a key signal for employees to interpret the intentions of the organization as well as to make inferences about which employee behaviors the organization values and rewards'' (Bowen \& Ostroff, 2004; Nishii et al., 2008; Ostroff \& Bowen, 2016, as cited in Nyfoudi et al., 2023, p.~541). When organizations create voice opportunities, they signal respect for employee expertise and psychological safety. This is particularly salient during organizational turbulence, where voice mechanisms may help employees as well as organizations.

The affective benefits of voice extend beyond implementation outcomes. Voice opportunities signal ``a work environment, in which employees feel safe to express themselves'' and create ``a Kantian type of environment, where the organization leaves `nobody alone in the crisis'\,'' \citep[p.~542]{nyfoudi2023}. This psychological safety and perceived organizational support enhance affective well-being, which research consistently links to improved job performance \citep{wright2000}. Importantly, this mechanism can operate even when specific suggestions are not implemented, as the act of soliciting and acknowledging input itself communicates respect and value. Voice provides workers with a sense of control and influence over their work environment, reducing frustration and enhancing commitment, independent of implementation.

\hyp{H3.}{The relationship between employee AI-voice and employee performance is positively mediated by employee affective well-being.}

\subsection{AI-Mediated Voice: Opportunities and Challenges}

AI-enabled listening systems may fundamentally reshape how employee voice translates into performance. Algorithmic tools can collect input at higher frequency, aggregate dispersed knowledge at scale, and systematically identify patterns across large volumes of employee feedback \citep{tambe2019}. By reducing managerial bandwidth constraints, AI may enhance the informational pathway of voice---accelerating the identification and routing of actionable improvement opportunities.

At the same time, voice research emphasizes that speaking up depends on psychological safety and perceived relational responsiveness \citep{morrison2014,sherf2019}. Traditional voice mechanisms embed informational exchange within interpersonal interaction, where managerial attention signals respect and inclusion. AI-mediated listening may alter this signal. Employees may perceive algorithmic processing as impersonal or opaque, potentially weakening the relational benefits associated with being heard. Emerging evidence suggests that digital and AI-enabled communication can shape psychological safety in distinct ways, depending on perceived anonymity, trust, and transparency \citep{griffith2025}.

These dynamics suggest that AI mediation may decouple the informational and affective functions of voice. While algorithmic systems may strengthen knowledge aggregation and implementation efficiency, they may simultaneously attenuate---or transform---the relational signals traditionally embedded in human listening. Whether AI ultimately substitutes for, complements, or reshapes human-mediated voice remains an open empirical question.

% ---------------------------------------------------------------------
\section{Research Design and Methodology}

\subsection{Empirical Setting}

The study will be conducted at a global strategy management consulting firm with about 30,000 employees, leveraging the weekly team pulse surveys administered to all consultants worldwide. This setting offers several advantages. First, professional services represent knowledge-intensive work where employee voice may be particularly valuable, as consultants possess specialized expertise and client-facing insights that managers cannot directly observe. Second, the weekly survey structure provides high-frequency longitudinal data, enabling analysis of voice dynamics over time. Third, the firm's experimentation with different voice mechanisms---including standard surveys, enhanced human-led options, and new AI-enabled channels---creates natural variation in the independent variable. The AI agents serve as the intervention.

\subsection{Dataset}

The dataset builds on the firm's Central European weekly Teaming Core Survey, which captures structured employee voice at the project (case) level across Germany and Austria, Switzerland, and Eastern Europe (Warsaw, Budapest, Prague). Importantly, surveys are not equivalent to cases: complex cases with multiple workstreams often generate several survey instances (submodules), and surveys are only activated once at least three team members are staffed. In a normal reference week, about 250 surveys are active, covering between 1,200 and 1,500 unique consultants (without double counting across submodules), of whom approximately 80\% are in human-led, coached settings supported by a teaming coach. The overall response rate in 2025 was exactly 60\%.

Survey items are measured on a four-point Likert scale ranging from 1 (``Agree'') to 4 (``Disagree'') and include validated well-being indicators (e.g., overall project experience, growth and learning, perceived value of contributions, psychological safety and team trust, workload sustainability, and non-case workload manageability). Two open-text fields---``What went particularly well?'' and ``What should the team consider doing differently?''---provide qualitative voice input; manual sampling over a sample month indicated that approximately 19\% of surveys contain at least one written comment. Coding the number of actual improvement suggestions in these fields yields the improvement-measure count.

Beyond survey responses, the dataset is embedded in a comprehensive performance and reporting infrastructure, including weekly case-level overviews, bottom and critical case detection, monthly account- and theme-clustered rankings, quarterly reporting, and annual account-level dashboards. Individual-level performance outcomes are measured using performance ratings and related evaluation metrics, enabling longitudinal linkage between voice exposure, improvement implementation, affective well-being, and subsequent performance. To establish this linkage, we rely on a persistent consultant-level identifier.

Overall, the database combines high-frequency behavioral, qualitative, and performance data, making it uniquely suited to test the proposed dual-mediation model of AI- and human-mediated employee voice.

\Needspace*{10\baselineskip}
\subsection{Intervention}

The core intervention manipulates the level and form of employee voice exposure:
\begin{enumerate}[nosep,leftmargin=*,label=\arabic*.]
  \item \textbf{Control condition:} no structured voice mechanism.
  \item \textbf{Human-led teaming coach:} human- and survey-based listening.
  \item \textbf{AI-enabled teaming agent:} agentic listening through a weekly conversation.
\end{enumerate}
Participants will be randomly assigned to conditions, with performance tracked over a three- to six-month period. This design allows causal inference regarding voice effects while comparing the relative efficacy of human-mediated versus AI-mediated listening.

\subsection{Measurement}

The firm's existing tracking system for voice episodes is similar to the ``formalized process to collect, track, and implement improvement-oriented issues'' described by Romney et al.\ \citeyearpar[p.~10]{romney2024}. We operationalize the measures as follows.

\subsubsection{Independent variable (employee voice)}
Voice will be operationalized through (1) the participation rate in teaming and surveys (participated/did not participate) and (2) the channel type (human-based or agent-based interaction).

\subsubsection{Mediator 1 (improvement measures)}
The variable will be operationalized through (1) the number of suggestions submitted per consultant (coded as a count); (2) the number of suggestions implemented (coded following Romney et al., 2024); and (3) the type of suggestion (coded as promotive or prohibitive, following Liang et al., 2012).

\subsubsection{Mediator 2 (affective well-being)}
Well-being will be measured using validated survey items embedded in the Teaming survey, including (1) WHO-5 Well-Being Index items \citep{topp2015} and (2) overall job/project satisfaction \citep[similar to][]{dolbier2005}.

\subsubsection{Dependent variable (employee performance)}
The primary performance metric is a consultant-level performance rating that combines perspectives from supervisors to client impressions. This is the core performance evaluation metric for consultants, measured on a standardized scale, and it incorporates multiple dimensions including client impact, analytical quality, teamwork, and leadership. Secondary performance indicators may include (1) core performance metrics and matrix placements; (2) peer and supervisor performance ratings; and (3) client ratings and impressions.

\subsection{Analytical Strategy}

\subsubsection{Mediation analysis}
We will employ structural equation modeling (SEM) to test the dual-mediation model, examining both direct and indirect effects of voice on performance through (1) process improvements and (2) affective well-being. Following \citet{hayes2018}, we will use bootstrapping procedures to estimate indirect effects and construct bias-corrected confidence intervals. Control variables will include consultant tenure, office location, practice area, and baseline performance ratings.

\subsubsection{Frequency effects}
Building on Nelson and Wilmers' \citeyearpar{nelson2025} finding that more intensive use of worker voice and the use of different voice formats are associated with higher productivity and earnings, we will analyze descriptively the usage of the AI voice systems.

% ---------------------------------------------------------------------
\section{Expected Contributions and Implications}

\subsection{Theoretical Contributions}

\begin{enumerate}[leftmargin=*,itemsep=4pt,label=\arabic*.]
\item \textbf{Extending voice theory to knowledge work.} Most empirical evidence on voice and productivity comes from blue-collar contexts \citep{nelson2025,wilmers2024}. By examining professional services, this study tests whether voice mechanisms generalize to knowledge-intensive work where employee expertise is less observable and more tacit.

\item \textbf{Disentangling process and affective mechanisms.} While prior research documents positive voice--performance relationships, the underlying mechanisms remain underexplored. By simultaneously testing process-based improvement and affective well-being pathways, this study provides a more nuanced understanding of how voice creates value. This addresses Romney et al.'s \citeyearpar{romney2024} call to ``recast prohibitive voice as input worth sharing because it initiates supervisor responsiveness that leads to the implementation of organizational improvements'' (p.~22), while also acknowledging the intrinsic value of being heard.

\item \textbf{Theorizing AI-mediated voice.} The study develops and tests theory regarding whether AI-enabled listening can replicate human-led voice effects. This extends emerging research on AI \citep{tambe2019} and addresses a critical gap identified by \citet{nelson2025}: whether multiple listening channels, including digital tools, can enhance productivity outcomes.

\item \textbf{Voice channel effectiveness.} By comparing traditional surveys, enhanced personalized mechanisms, and AI-enabled tools, the study contributes to understanding which voice channels are most effective under what conditions, responding to Starzyk and Bauer's \citeyearpar{starzyk2025} analysis of how voice opportunities are shaped by organizational structures.
\end{enumerate}

\subsection{Practical Contributions}

\begin{enumerate}[leftmargin=*,itemsep=4pt,label=\arabic*.]
\item \textbf{Guidance for voice system design.} Organizations investing in employee listening platforms need evidence-based guidance on channel selection, frequency, and implementation. This study will provide actionable insights on whether AI tools can cost-effectively scale voice mechanisms while maintaining or enhancing their benefits.

\item \textbf{Technology investment decisions.} With the proliferation of AI-enabled HR technologies, firms need evidence on their effectiveness. This study will help organizations determine whether investments in agentic listening tools yield returns comparable to traditional approaches.

\item \textbf{Employee well-being interventions.} If the affective pathway proves significant, voice mechanisms represent a scalable intervention for enhancing employee well-being---a critical concern in high-intensity professional services contexts.
\end{enumerate}

% ---------------------------------------------------------------------
\section{Limitations and Next Steps}

This study has several limitations. First, the single-firm setting may limit generalizability, and replication across industries and organizational contexts is warranted. Second, although the experimental design enables causal inference regarding voice exposure, differences between AI- and human-mediated conditions may reflect variation in implementation quality, necessitating careful fidelity checks. Third, performance is measured using organizational performance ratings, which may incorporate evaluative bias or political dynamics; triangulation with alternative indicators such as profitability or client retention would strengthen inference.

Future research should examine boundary conditions and potential unintended consequences of voice. For example, voice effectiveness may depend on broader HR-system strength and crisis-related practices \citep{nyfoudi2023}. Moreover, prohibitive voice can impose cognitive and emotional burdens, potentially eliciting exhaustion or defensive reactions \citep{romney2024}. Understanding when voice creates strain rather than value remains important.

% ---------------------------------------------------------------------
\section{Conclusion}

This research proposal investigates a fundamental question at the intersection of employee voice, AI-enabled HRM, and performance management: can AI-mediated listening mechanisms enhance the productivity benefits of traditional employee voice? By examining dual mediating pathways---process improvement and affective well-being---the study aims to disentangle the mechanisms through which voice creates value.

Theoretically, the study will advance understanding of voice mechanisms, test the generalizability of manufacturing-based findings to professional services, and develop new theory regarding AI-mediated voice effectiveness. Practically, the research will provide evidence-based guidance for organizations seeking to enhance employee performance and well-being through voice mechanisms, while navigating the opportunities and risks of AI-enabled listening.

In an era where ``listening to workers pays off in productivity'' \citep{wilmers2024}, understanding how to listen most effectively---and whether AI can serve as an effective listener---represents a critical frontier for organizational research and practice.

% ---------------------------------------------------------------------
\renewcommand{\refname}{References}

\end{document}